\PassOptionsToPackage{unicode}{hyperref}
\PassOptionsToPackage{hyphens}{url}
\PassOptionsToPackage{dvipsnames,svgnames*,x11names*}{xcolor}
\documentclass[
  12 pt,
]{paper}
\usepackage{lmodern}
\usepackage{amssymb,amsmath}
\usepackage{ifxetex,ifluatex}
\ifnum 0\ifxetex 1\fi\ifluatex 1\fi=0 
  \usepackage[T1]{fontenc}
  \usepackage[utf8]{inputenc}
  \usepackage{textcomp} 
\else 
  \usepackage{unicode-math}
  \defaultfontfeatures{Scale=MatchLowercase}
  \defaultfontfeatures[\rmfamily]{Ligatures=TeX,Scale=1}
\fi
\IfFileExists{upquote.sty}{\usepackage{upquote}}{}
\IfFileExists{microtype.sty}{
  \usepackage[]{microtype}
  \UseMicrotypeSet[protrusion]{basicmath} 
}{}
\makeatletter
\@ifundefined{KOMAClassName}{
  \IfFileExists{parskip.sty}{%
    \usepackage{parskip}
  }{
    \setlength{\parindent}{0pt}
    \setlength{\parskip}{6pt plus 2pt minus 1pt}}
}{
  \KOMAoptions{parskip=half}}
\makeatother
\usepackage{xcolor}
\IfFileExists{xurl.sty}{\usepackage{xurl}}{} 
\IfFileExists{bookmark.sty}{\usepackage{bookmark}}{\usepackage{hyperref}}
\hypersetup{
  pdftitle={Opinionated practices for teaching reproducibility: motivation, guided instruction and practice},
  pdfauthor={Joel Ostblom Department of Computer Science, University of British Columbia; Tiffany Timbers Department of Statistics, University of British Columbia; Corresponding author tiffany.timbers@stat.ubc.ca},
  colorlinks=true,
  linkcolor=blue,
  filecolor=Maroon,
  citecolor=Blue,
  urlcolor=Blue,
  pdfcreator={LaTeX via pandoc}}
\usepackage{longtable,booktabs}
\usepackage{etoolbox}
\makeatletter
\patchcmd\longtable{\par}{\if@noskipsec\mbox{}\fi\par}{}{}
\makeatother
\IfFileExists{footnotehyper.sty}{\usepackage{footnotehyper}}{\usepackage{footnote}}
\makesavenoteenv{longtable}
\usepackage{graphicx}
\makeatletter
\def\maxwidth{\ifdim\Gin@nat@width>\linewidth\linewidth\else\Gin@nat@width\fi}
\def\maxheight{\ifdim\Gin@nat@height>\textheight\textheight\else\Gin@nat@height\fi}
\makeatother
\setkeys{Gin}{width=\maxwidth,height=\maxheight,keepaspectratio}
\makeatletter
\def\fps@figure{htbp}
\makeatother
\providecommand{\tightlist}{%
  \setlength{\itemsep}{0pt}\setlength{\parskip}{0pt}}
\usepackage{setspace}
\usepackage[style=authoryear,]{biblatex}
\title{Opinionated practices for teaching reproducibility: motivation, guided instruction and practice}
\author{Joel Ostblom \small Department of Computer Science, University of British Columbia \and Tiffany Timbers \small Department of Statistics, University of British Columbia \and \footnotesize Corresponding author \href{mailto:tiffany.timbers@stat.ubc.ca}{\nolinkurl{tiffany.timbers@stat.ubc.ca}}}
\date{}

\begin{document}
\maketitle
\begin{abstract}
In the data science courses at the University of British Columbia,
we define data science as the study, development
and practice of reproducible
and auditable processes to obtain insight from data.
While reproducibility is core to our definition,
most data science learners enter the field with other aspects of data science in mind,
for example predictive modelling, which is often one of the most interesting topic to novices.
This fact, along with the highly technical nature
of the industry standard reproducibility tools currently employed in data science,
present out-of-the gate challenges in teaching reproducibility in the data science classroom.
Put simply, students are not as intrinsically motivated to learn this topic,
and it is not an easy one for them to learn. What can a data science educator do?
Over several iterations of teaching courses focused on reproducible data science tools and workflows,
we have found that providing extra motivation, guided instruction
and lots of practice are key to effectively teaching this challenging, yet important subject.
Here we present examples of how we deeply motivate, effectively guide
and provide ample practice opportunities to data science students
to effectively engage them in learning about this topic.
\end{abstract}


\newpage

\hypertarget{introduction}{%
\subsection{Introduction}\label{introduction}}

In the graduate and undergraduate data science courses that we teach
at the University of British Columbia (UBC),
we define data science as the study, development, and practice of
reproducible and auditable processes to extract insight from data.
Using this definition requires that we also define
what is meant by a reproducible and auditable analysis.
Although the specific definition of reproducibility
varies between research domains
\autocite{committeeonreproducibilityandreplicabilityinscience_2019},
in our data science courses
we have chosen to embrace the National Academy of Sciences definition of reproducible analysis,
which is reaching the same result given the same input, computational methods,
and conditions \autocite*{national2019reproducibility}.
For auditable or transparent analysis,
we follow how it has been defined by others \autocite{parker2017opinionated,ram2013git},
which is that there should be a readable record of the steps used to carry out the analysis
(i.e., computer code)
as well as a record of how the analysis methods evolved
(i.e., a version controlled project history).
This history is important for recording how and why
decisions to use one method or another were made,
among other things.

The reason we embrace this definition of data science,
is that we believe data science work should both bring insight
(e.g., answer an important research question)
and employ reproducible and auditable methods
so that trustworthy results and data products can be created.
Results and data products can be generated
without reproducible and auditable methods,
however,
when they are built this way there is less confidence
in how the results or products were created.
We believe this stems from non-reproducible and non-auditable analyses:

\begin{enumerate}
\def\labelenumi{\arabic{enumi}.}
\tightlist
\item
  lacking evidence that the results or product could be regenerated given the same input, computational methods, and conditions
\item
  lacking evidence of the steps taken during creation
\item
  having an incomplete record of how and why analysis decisions were made
\end{enumerate}

In addition to contributing to the trustworthiness of data science work,
employing reproducible and auditable methods and workflows
bring additional benefits to data scientists,
such as more effective collaboration.
Data science is an inherently collaborative discipline,
and adhering to reproducible and auditable data science methods
greatly facilitates the act of collaborating in many contexts,
further emphasizing the importance of learning this skill well.

Although the many benefits of reproducible and auditable analyses discussed above
may make this seem like an exciting topic for incoming students,
the experience when entering a classroom of curious data scientists in training
is quite often the opposite.
Students are usually keen to learn about data science
but what they are most often excited about is the second part of
its definition: extracting insights from the data.
Commonly, students are not even aware of the reproducible
and auditable processes of data science,
and when they first hear about them,
they tend to regard them as an inconvenient means to an end
rather than an important skill to master.
This outlook is likely at least in part motivated
by the fact that these processes
do not directly lead to novel insights
in the same way as a predictive model might,
which is what many students have in mind when they envision the work of a data scientist.
This negative predisposition creates another barrier to overcome
when teaching the reproducible and auditable aspects of data science.

An additional pedagogical challenge
is that the tools that we use for reproducibility
are not necessarily smooth and easy to learn,
but often have a steep learning curve.
Over our five years of teaching these topics at UBC
we have found three pedagogical strategies
that are particularly effective for teaching reproducibility successfully:

\begin{enumerate}
\def\labelenumi{\arabic{enumi}.}
\tightlist
\item
  placing extra emphasis on motivation
\item
  guided instruction
\item
  lots of practice
\end{enumerate}

In this paper,
we will discuss why we believe each of these are important,
and provide examples of how to incorporate these in your teaching.

\hypertarget{academic-setting-and-background}{%
\subsubsection{Academic setting and background}\label{academic-setting-and-background}}

The specific teaching pedagogies we present in this paper
have been tested over five years,
in classrooms of 20-200 learners from varying educational backgrounds,
including undergraduate students taking their first data science course
and graduate students looking to specialize in data science.
Most of the courses where we teach reproducibility
are part of UBC's professional Master of Data Science program.
This course-based program consists of 24 one-credit courses
taught over an eight month period,
which culminates in a six-credit-capstone course.
As a prerequisite for this program
students are generally required to have taken at least one course in each of
calculus or linear algebra, computer programming, and introductory statistics or probability.
Most of these graduate students
have several years of professional work experience (or equivalent),
however our cohorts also include both recent graduates from Bachelor's degrees
as well as learners from senior positions in industry who are
looking to bend their career trajectories towards data science.
From in-class polls,
we have learned that these graduate students
rarely have any prior experience with reproducible workflows.

The teaching team for our reproducibility courses
include a mix of postdoctoral teaching and learning fellows,
lecturers, and assistant professors of teaching
from the departments of Statistics and Computer Science.
A handful of the core members of the program's teaching team
have been with the program since its inception in 2016,
and the remainder of the team has been with the program for 1-3 years.
Members of the teaching team developed their data science expertise
through graduate and postdoctoral studies in a diverse set of research domains,
including computational linguistics, ecotoxicology, learning analytics,
machine learning, neuroscience, physics, statistics, and stem cell engineering.
Our pedagogies are influenced not only by our formal training at UBC,
but also by initiatives such as
The Carpentries \autocite{__aq},
RStudio education \autocite{__ap},
Teaching Tech Together \autocite{wilson_wilson_2018},
and the Instructional Skills Workshop \autocite{day_day_2005}.

Since the launch of the program six years ago,
our teaching strategies have evolved
in response to student feedback on our teaching practices
obtained through:
\emph{1)} student-led surveys and face-to-face meetings,
\emph{2)} instructor-led course surveys,
\emph{3)} formal university course evaluations,
and \emph{4)} surveys of alumni of the program.
An example of a direct improvement in the program
from the collection and implementation of student feedback
was the addition of lecture worksheets with simpler and often autograded questions.
This was done to address common feedback
we received across the program's curriculum
about the difficulty-level of lab homework assessments.
Specifically, students reported that they perceived
there to be too large of a jump in difficulty
from what was learned in the lecture
to what was expected from them on the lab homework assessments.
Implementing lecture worksheets in several of our courses
decreased the amount of feedback we received
about the difference in difficulty
between the lectures and assessments in those courses.
We hypothesize this is particularly true for students' who
have less interest, experience, or skills in the area of reproducibility,
and we have been incorporating these worksheets
across more and more courses in the program to continue to address this.

One of the key reproducibility-related takeaways from our student feedback
is that practice matters and that more is better.
In the capstone projects from the first year of the program,
instructors observed that students could easily do the things
we gave them lots of practice on
(e.g., basic version control
and working linearly on the default branch),
but did not attempt or struggled with other reproducibility concepts,
skills and workflows where they were provided formal instructions
but less practice
(e.g., more advanced version control topics such as branching, pull-requests, and
advanced usage of reproducible reports and presentations).
In response to this,
we implemented project-based course-work for several reproducibility related topics
across the program so that students had more opportunity to practice these.
The instructors observed that the capstone projects in the years following this change
had less technical debt and included more, and higher quality reproducible practices.
Encouragingly,
our capstone partners also noticed this improvement
in the quality of the student projects,
which was reflected in the feedback we collected from them.

\hypertarget{placing-extra-emphasis-on-motivation}{%
\subsection{Placing extra emphasis on motivation}\label{placing-extra-emphasis-on-motivation}}

Why do we need extra motivation when teaching reproducibility,
compared to some other data science topics, such as machine learning?
Our experience is that this stems from students' lack of intrinsic excitement
or motivation for the topic of reproducibility,
that they have little prior knowledge on this topic,
and that reproducibility concepts and in particular tools are challenging to learn.

\begin{figure}

{\centering \includegraphics[width=0.3\linewidth]{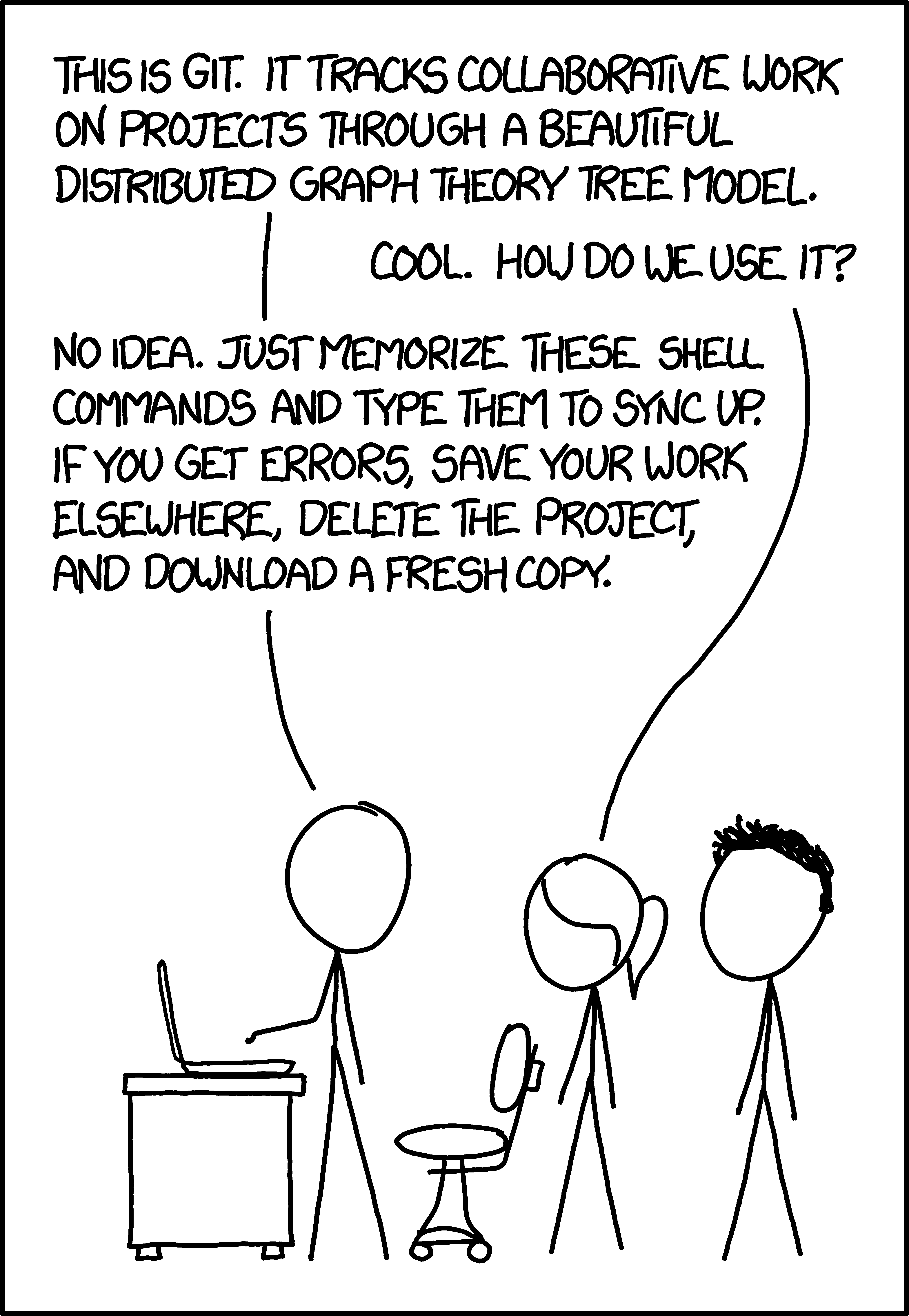} 

}

\caption{Infamous comic from \href{https://xkcd.com}{xkcd.com} that highlights the difficulty of learning and using the version control software Git.}\label{fig:git-is-hard-xkcd}
\end{figure}

One example is that the most commonly used version control software, Git,
is notorious for being difficult to learn (Figure \ref{fig:git-is-hard-xkcd}).
There are many anecdotes attesting that most people
do not learn Git deeply
and instead work by trying a variety of commands until they find some
that work. This can lead to learners getting themselves into challenging or
perplexing version control situations, with difficult to interpret outputs
(e.g., ``You are in `detached HEAD' state. You can look around, make experimental changes and commit them, and you can discard any commits you make in this state without impacting any branches by performing another checkout'').
Sometimes these situations are so difficult to get out of, that even professional
data scientists and data science educators
recommend the practice of ``burning it all down'' and starting from scratch \autocite{bryan_bryan_2021} -
which really defeats many of the purposes of version control.

\begin{figure}

{\centering \includegraphics[width=1\linewidth]{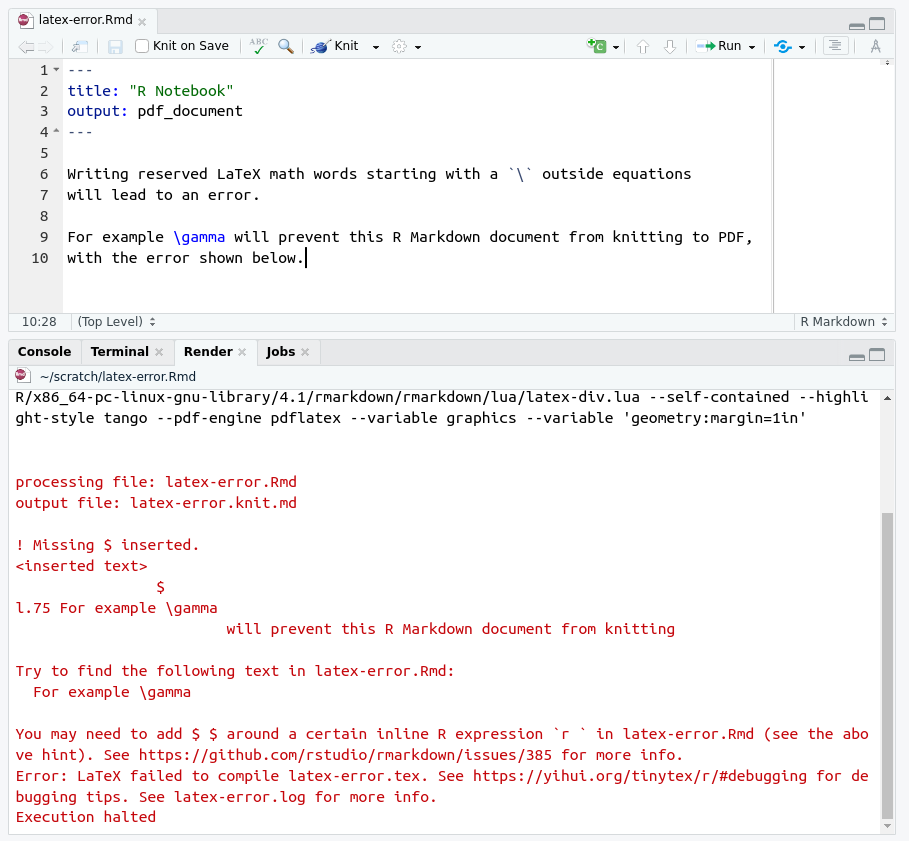} 

}

\caption{LaTeX errors are often cryptic to new learners.}\label{fig:latex-errors}
\end{figure}

Another example is R Markdown,
which is an implementation of literate code documents \autocite{knuth1984literate},
that are useful for generating reproducible reports.
Many aspects of R Markdown are quite user friendly,
however rendering the source R Markdown document to PDF depends on LaTeX.
If users make formatting errors that impact LaTeX's job in the rendering
the resulting error messages can be cryptic
and are often not clear to learners about where the error is coming from
or how to resolve it.
An example is the error message from including a ``\textbackslash{}'' character with a reserved
LaTeX word outside of a mathematical equation,
which is shown in Figure \ref{fig:latex-errors}.
This error message is interpretable by intermediate and experienced
R Markdown users, but is challenging for novice learners to parse.

Yet another example of a popular reproducibility tool that is challenging to learn is
Docker.
Docker is a containerization tool that extends computational environment management beyond
the programming language and package dependencies of your analysis workflow,
and instead creates a versioned copy of your entire computational environment
including any additional software and operating system dependencies that your analysis may depend on.
This means that you can share an exact copy of your computational environment
with your collaborators, even if you work on different operating systems.
Docker is an especially challenging tool to
teach and motivate students to learn
because it is so different from writing code for analysis.
This largely stems from software installation being a time consuming process,
which means that writing and debugging code to automate installation
is a slow and painstaking process.
This issue is further exacerbated
by the fact that this is not an exciting part of a data analysis,
as things already work on the student's own computer.
Together,
these issues make it difficult to convince students
to put effort into learning and using this tool.

So what can we do to motivate learning reproducibility concepts and tools in our
classrooms? We have found the following three strategies helpful:

\begin{enumerate}
\def\labelenumi{\arabic{enumi}.}
\tightlist
\item
  Tell stories from the trenches
\item
  Study cases of failures with real world consequences
\item
  Let them fail (in a controlled manner)
\end{enumerate}

\hypertarget{telling-stories-from-the-trenches}{%
\subsubsection{Telling stories from the trenches}\label{telling-stories-from-the-trenches}}

One successful strategy that we have used is telling stories from the trenches.
The instructors who teach these courses at UBC usually have had some experience
performing data analyses in their Ph.D.'s or Postdoctoral studies,
or are still currently engaged in research where they do this.
Through these lived experiences
of learning reproducibility tools
and applying them to our research,
we have both made our own mistakes
and witnessed those of our collaborators,
and thus we can share these experiences with our students.
In the Master of Data Science program,
many of the students have work experience involving data in the past,
so they also have stories from the trenches.
In a classroom with such students,
you can carry out think-pair-share exercises around these stories
and get the students to talk about their stories as well as hear yours.
Think-pair-share is a cooperative discussion exercise
where students first \emph{think} about a prompt on their own,
then \emph{pair} up to discuss their response with a few classmates,
and finally \emph{share} their thinking with the rest of the class afterwards \autocite{lyman_lyman_1981}.
We think this pedagogy is particularly effective in this context
as it allows students to share their past mistakes in a smaller group first.
There they can hear their peers' similar stories,
and they will observe that reproducibility challenges are widespread
and virtually everyone who has previously worked with data has experienced these.
This can support them in feeling more confident and secure
before choosing whether to share their story with the entire class.

A specific example of how we instruct students to run a think-pair-share exercise
on the topic of reproducibility follows:

\begin{enumerate}
\def\labelenumi{\arabic{enumi}.}
\tightlist
\item
  Students are prompted to think about a non-reproducible or non-auditable workflow
  they have used before at work, on a personal project, or in course work,
  and that negatively impacted their work somehow
  (make sure to include this in the story).
\item
  The instructor shares their own example to get the students started.
\item
  Students share their story
  and how it negatively impacted their workflow
  with the person beside them.
\item
  Students are then asked to share their story
  in a course forum or collaborative note-taking document
  that other students could read
  (in smaller classes all groups could share one of their stories aloud with the class).
\end{enumerate}

In addition to sharing some consequences of non-reproducible analysis with students
without them having to experience the negative impact themselves
over a long and drawn out process of self-discovery,
these stories come from people the students know
(their instructors and peers).
This highlights how common and easy it is for such mistakes to happen
to anyone practicing data analysis,
unless care is taken to follow reproducible workflows and practices.
This sharing of stories
helps make the practice of reproducibility
seem more relatable and applicable to the students' own data analysis
(which at this stage in their career, may not end up in a published paper).

\hypertarget{examples-of-stories-from-the-trenches}{%
\paragraph{Examples of stories from the trenches}\label{examples-of-stories-from-the-trenches}}

\begin{quote}
As a Masters student, I started to use R to do my statistical analysis. I obtained the results I needed from running my code in the R console and copying the results into the word document that was my manuscript. Six months later we were working on revisions requested by the reviewers and I could not remember which version of the code I ran to get my results. I eventually figured it out through much trial and error, but the process was inefficient and very stressful.

-- Tiffany Timbers
\end{quote}

\begin{quote}
I was involved in a project where we used version control for the code, but didn't keep track of which input/output was analyzed/produced with which version of the code base. This happened because the data was too big for a simple solution like GitHub, and I was under time pressure to produce results so I didn't prioritize looking into an appropriate solution. When I returned to this project after a long absence, I could not easily combine the outputs from my earlier analysis with the newly generated ones, and instead had to re-analyze all the data with the latest version of the code to reduce the chance for issues from using conflicting code bases. This was \emph{very} time-consuming. As often is the case in projects where code is only seen as a means to an end and not part of the final product, there was also no time dedicated to write tests for the code in this project, so there was no guarantee that there were not unintended side effects introduced when new changes were made.

-- Joel Ostblom
\end{quote}

\hypertarget{study-cases-of-failures-with-significant-real-world-consequences}{%
\subsubsection{Study cases of failures with significant real world consequences}\label{study-cases-of-failures-with-significant-real-world-consequences}}

A second way to create motivation
is through using case studies of irreproducible data analyses
that have had significant real world consequences.
Such case studies can be used to illustrate the importance
and impact of reproducible data analyses.
Although we are currently in the process of building case-based teaching
into our reproducibility courses
and currently only have limited experience with this pedagogy,
we think the idea has strong merit
and wish to present it here.
Notably,
case-based teaching has been widely used
in business, law, and medical education for many years \autocite{carlson1995beyond,garvin_garvin_2003,bonney_bonney_2015},
where it has been shown to motivate students
to participate in class activities to a higher extent,
which boosts student assessment performance \autocite{flynn2001influence,yadav2007teaching}.
The benefits of case studies have also been reported on in STEM fields \autocite{yadav2007teaching,bonney_bonney_2015}
and we believe that this pedagogical strategy
may be particularly important for teaching reproducibility,
since the impact and significance of the consequences
of not using reproducible practices are not obvious to novices.
By presenting case studies
where failure to adhering to reproducible practices
have led to costly mistakes,
we aim to give learners a chance to directly appreciate
the connection between the lack of reproducible workflows
and the downstream consequences.

While there are many articles outlining recommendations
on which reproducible practices to adhere to,
case studies of failures are not as frequent in the literature.
We think this partly stems from the fact
that such errors are often only discovered internally and never reported,
and that there are few incentives for people
to spend their time performing proper forensic informatic analysis on the work of others.
Even so,
there are several such examples reported in the literature,
one of the most striking led to putting patients at risk
in incorrectly administered clinical trials
which we will outline in the next paragraph \autocite{carlson_carlson_2012}.

These clinical trials took place at Duke in 2006
and involved 110 cancer patients
hoping that using personalized gene signatures
would identify which treatments were more effective for individuals \autocite{carlson_carlson_2012}.
The series of scientific papers that formed the basis of these trials
were all published in highly regarded ``high-impact'' journals,
however they also raised some concerns among researchers in the field \autocite{carlson_carlson_2012}.
When put under a thorough independent review,
these papers were found to contain multiple errors,
several related to the use of non-reproducible tools
and workflows \autocite{baggerly2009deriving}.
In the review analysis it was highlighted that the most common problems were simple
and included mistakes such as: \emph{1)} ``off-by-one'' errors
where a cell might have been inadvertently deleted in a spreadsheet
leading to a shift of all remaining values,
\emph{2)} labelling mix-ups where the treated and not treated groups
were assigned labels 0 and 1 instead of meaningful names
which can lead to confusion as to which is the treated and control group,
and \emph{3)} poor documentation practices leading to lack of transparency
which makes it harder and more time-consuming to identify errors
(both for the original authors and the reviewers).
These clinical trials were eventually terminated about four years after they started,
around 25 papers related to these trials were retracted,
and the lead investigators were put under investigation for malpractice.
This example highlights the enormous cost
associated with not adhering to reproducible practices
and having workflows that are opaque and hard to review.

As illustrated in the case we have highlighted above,
reproducibility errors are not isolated to novices performing data analyses.
They also occur in analyses performed by seasoned professionals
and can have substantial real-world consequences.
When selecting failures to share with students,
it can be beneficial to focus on
those that students can easily identify with,
those that have had a big impact,
and those that come from well-known researchers/companies.
This is to illustrate that these errors can happen to anyone
and emphasize that we are not sharing them to cast blame,
but to highlight the importance of reproducible practices on all levels.
Another example of an impactful reproducibility mistake from a well-known research team
is the misreporting of the relationship between public debt and gross domestic product (GDP) growth,
which was published in what was regarded as a seminal paper at the time \autocite{pollin_herndon_2013,borweinjon_bailey_2013}.
These findings came from prominent economists
and were used as motivation for political decision making,
until a few years later when it was discovered that an error had been made
when selecting the range of cells to be included in one of the Excel formulas,
which exaggerated the conclusions in the paper
\autocite{pollin_herndon_2013,borweinjon_bailey_2013}.
Other examples of case studies that could be used to illustrate the significant
real world consequences of reproducibility failures in data analyses
are listed in Table 1.

\hypertarget{letting-them-fail-in-a-controlled-manner}{%
\subsubsection{Letting them fail (in a controlled manner)}\label{letting-them-fail-in-a-controlled-manner}}

Many instructors (including the authors of this work)
have themselves experienced failure in graduate school
and during their postdoctoral research in regards to reproducibility,
which negatively impacted their work.
While these experiences motivate teaching and using
reproducibility concepts and tools for instructors,
most undergraduates and new graduate students cannot draw on similar professional experiences.
Rather than letting new students live through the full perils of irreproducible research,
we can set up controlled scenarios to expose them to these downsides
in a controlled, accelerated manner
while still embodying much of the same motivational benefits.

One way we have done this is providing students an analysis that is not
reproducible
(i.e., it depends on rare/obscure software packages
or specific package versions),
and thus will likely fail on someone else's machine.
We then ask them to try to run it,
and if they cannot run it, we ask them to fix the code
or install missing software so they can.
Then we provide them the same analysis that has been made reproducible,
through the use of shareable compute environments (i.e., \texttt{renv}, \texttt{conda}, or \texttt{Docker}),
and instructions on how to use the shared compute environment.
The students then experience running the same analysis on their machines
without any change of code or software installation.
Under these controlled circumstances
failure and frustration can have a positive impact on students overall learning
as they experience the many benefits of reproducibility first hand.
This exercise usually only takes around 20-30 min
and helps provide motivation
to endure the steep learning curves of reproducibility concepts and tools.
We teach this in our data science workflows course,
which occurs in the first third of the Master of Data Science program.
When we have implemented this activity
we have run it as an in-class exercise
that is not counted toward the course grades.
We do however use polls to assess where the students are at
in regards to completing the activity,
so that we can allot an appropriate amount of time for each group of learners.

The example discussed above was inspired by Jenny Bryan's teaching of STAT 545 at UBC.
In her version of this task, she pairs students up and asks them to
run each other's code projects that they have been working on in her course.
Most usually fail to be able to do this on the first try
for the same reasons discussed above.
We adapted her teaching method so that students experience this
in a more limited, controlled manner
by picking an obscure package (e.g., the R-package cowsay \autocite{cowsay_2020})
that none of the students should have installed,
so that they will all have trouble running the project without reproducibility tools.
We then provide them with a project using reproducibility
tools so that they can have success and contrast the two experiences.
This has allowed us to scale this exercise to larger classes
with more homogeneous student experiences.

\hypertarget{guided-instruction}{%
\subsection{Guided instruction}\label{guided-instruction}}

In our teaching
we primarily seek to facilitate students' active learning
and encourage them to take initiative and responsibility for their own learning experience.
Here, we suggest that guided instruction is helpful
to set students off on the right path as they take an active role in their own learning,
particularly when teaching reproducibility.
From our experience, reproducibility is not something that most people
figure out on their own,
and if they do, it is an inefficient time-consuming process.
We hypothesize that this could be at least partially explained
by the fact that reproducibility practices
borrow knowledge, workflows, and tools from software engineering
and repurpose them for data science.
Thus, much of the getting up and getting started with reproducibility
has a lot of assumed knowledge behind it,
and at present there are not many clear and easy on-ramps for learners who
do not have a software engineering background.
Part of this may stem from the field being still fairly new
and not-yet as widely embraced as we might hope.
This means that there is not a lot of culture around using reproducibility tools in data science
and statistics, and it is not yet as obvious how to get started.

Furthermore, similar to why we need extra motivation,
the challenge of learning to use the tools due to their steep learning curves
suggests that having some guided instruction is beneficial to learners.
These points are well stated in a
\href{https://simplystatistics.org/2018/12/11/the-role-of-theory-in-data-analysis/}{blog post}
and \href{https://leanpub.com/dataanalysisessays}{essay}
on \emph{The Role of Theory in Data Analysis} by Roger Peng \autocite*{peng2020essays}:

\begin{quote}
There is no need for a new data analyst to learn about reproducibility
``from experience''.
We don't need to lead a junior data analysis down a months-long
winding path of non-reproducible analysis until they are finally bitten
by the non-reproducibility bug (and ``therefore learn their lesson'').
We can just tell them

``In the past, we've found it useful to make our data analysis reproducible
Here's a workflow to guide you in your own analysis.''

Within that one statement, we can ``compress'' over 20 years of experience.
\end{quote}

While we believe that there are pedagogical advantages
to letting students fail briefly in a controlled manner
(as elaborated on in the previous section),
we agree with the statement above in the sense that it is important
for educators to present the best practices
that the reproducibility community has arrived on to date
and explicitly show learners how to use these.
If we instead relied solely on students learning reproducibility through their own mistakes
we would set them up for a frustrating and time-consuming learning experience.

At UBC,
we primarily employ guided instructions through three pedagogical strategies:

\begin{enumerate}
\def\labelenumi{\arabic{enumi}.}
\tightlist
\item
  Live demonstration
\item
  Pre-lecture activities
\item
  Worksheets
\end{enumerate}

\hypertarget{live-demonstration}{%
\subsubsection{Live demonstration}\label{live-demonstration}}

In data science programming classes it is becoming more common to use demonstration
to show how to code in R and Python.
This is referred to in the data science literature as live coding
\autocite{raj2018role,nederbragt2020ten}.
We have observed that a similar pedagogy of live demonstration
works well for reproducibility tools,
including R Markdown \autocite{xie2021dynamic} or Jupyter notebooks \autocite{kluyver2016jupyter} for reproducible reports,
using version control with Git and GitHub,
and using tools like \texttt{renv} \autocite{kusheyrenv}, \texttt{conda} \autocite{anaconda}
and Docker \autocite{merkel2014docker} to create reproducible and shareable computational environments.
We believe that live demonstration makes it more obvious to the students
how to use these tools in practice,
and facilitates lateral knowledge transfer
where learners absorb additional material by observing \emph{how} we work,
which would not have been possible from learning about these
concepts and tools in a traditional lecture that uses a slide deck
to present new knowledge and concepts.
Additionally, when you make mistakes as an instructor in these live demonstrations,
it humanizes the reality of working with these tools that are
somewhat challenging even for experts,
and intentional mistakes can provide opportunities to spend more time on
that area of the topic and explain the gotchas of a common mistake,
and how to fix it \autocite{wilson_wilson_2018}.

A word of caution with live demonstration when teaching reproducibility;
because teaching these workflows and tools often involves
the demonstrations of graphical user interfaces
and tools that come from software engineering,
the tech stack moves very fast.
This means that each semester we teach these tools,
we need to test drive the materials before we share them with the students
to see if something has changed - as often it has.
A relatively recent example of this is from fall 2020,
when GitHub decided to change the name of their default branch
from master to main (rightfully so) \autocite{__ak}.
This change broke several of our teaching demonstrations,
guided worksheets, and lab homework.
It also caused parts of our lecture notes on this topic to have to be rewritten.
These rewrites happen less frequently when teaching
fundamental data science concepts,
as that part of the data science software stack has now become fairly stable.
For example,
teaching dataframes as the basic structure for encoding data for analysis
has remained constant since the inception of the program
and is unlikely to change any time soon.
In general,
changes in novice courses are often on details
such as new syntax or new naming conventions,
whereas changes in more advanced topics
can involve rewriting the entire coding component of a course
because there has been a notable shift in that domain
and there is now a more effective package or approach to teach.

For the reasons argued above, we believe that it is integral
to use guided instruction when teaching reproducibility,
however, this should be done with the awareness and the acceptance
that these kinds of changes are going to happen relatively frequently.
Which means reproducibility instructors are going to have to
update or make new live demonstrations, and other teaching resources,
each year.
Without this, the course resources will quickly fall out of usefulness.
There may come a time when these tools also stabilize,
but the authors of this manuscript anticipate
that to lie many years in the future.

\hypertarget{pre-lecture-activities}{%
\subsubsection{Pre-lecture activities}\label{pre-lecture-activities}}

Although live demonstration is important for the reasons outlined above,
it is critical that it does not dominate the time spent in the classroom,
so that students have ample time to engage in active learning activities.
There is evidence that active learning can increase student performance,
at least on summative assessments \autocite{freeman2014active}.
To ensure that students have sufficient background knowledge
to start, we complement our live demonstration with assigned pre-lecture activities.
These can consist of material that we have created ourselves or external resources
and is usually in the form of reading material or videos.
Encouraging students to learn the basics before class
allows us to have more meaningful live demonstrations
and sets students up for a more effective learning experience
by spacing out their exposure to the course material.
To assess whether students are understanding the material in the pre-readings
(and what we are covering during class),
we use in-class polls,
which allow us to adjust and spend more time on concepts
that had a lower percentage of correct answers.

\hypertarget{worksheets}{%
\subsubsection{Worksheets}\label{worksheets}}

After guiding students through the fundamentals
through live demonstration and pre-lecture activities,
we challenge them to take a more active learning role
by solving worksheet problems on their own in the classroom.
This activity occupies most of the in-class time,
so that students can engage actively with the material
in an environment where they can easily be supported
by their peers and the teaching team,
before working on the homework assignments on their own.
Worksheets are low stakes assessments that
provide students with many short problems on which to practice and receive feedback.
In data science,
this works well in literate code documents
(either Jupyter notebooks or R Markdown)
that have automated tests in them to provide feedback.
Two tools that we have used for this are nbgrader \autocite{blank2019nbgrader}
and otter-grader \autocite{otter}.
Compared to in-class exercises,
worksheets give learners an additional chance to actively engage with the material
while still providing structured exercises
focused on key learning outcomes.
Worksheets are also key for giving students ample opportunity to practice,
a topic we discuss in the last section of this article.

\hypertarget{example-lesson-using-guided-instruction}{%
\subsubsection{Example lesson using guided instruction}\label{example-lesson-using-guided-instruction}}

Here we provide an example of how we use guided instruction to teach version
control in our first year introduction to data science course at UBC, DSCI 100.
In this course, we take a three-pronged approach for guided instruction.

First, we provide students an assigned textbook reading
for them to review before class.
For this particular topic, it is an \href{https://datasciencebook.ca/Getting-started-with-version-control.html}{introductory chapter
on collaboration with version control}.
When the students arrive in class,
we then do a live demonstration,
where they watch us use the GitHub website,
and the Jupyter Git graphical user interface to add, commit,
push and pull changes.
A recording of one of these live demonstrations
is available on YouTube at this \href{https://youtu.be/attPo4zEElU}{link}.
Finally we ask the students to work through a
guided worksheet, which is a Jupyter notebook with narration, questions and automated software tests to give automated feedback about their answers.
The worksheet asks the students to perform
the same task that we just demonstrated,
and asks them questions along the way
to test their understanding of the reproducibility concepts
related to the skills and tools they are practicing.
For example, in the version control worksheet, we ask questions to assess
if they understand what the purpose of adding something to the Git staging area is,
how adding differs from committing in Git,
and where the work goes when it is pushed to a remote repository.
A version of \href{https://github.com/UBC-DSCI/data-science-a-first-intro-worksheets/blob/main/worksheet_version_control/worksheet_version_control.ipynb}{this worksheet can be accessed here}
as part of the actively maintained repository of worksheets that act as a companion to the
textbook we use for DSCI 100: \emph{Data Science: A First Introduction} \autocite{lee_timbers_2022}.

The challenge or limitation with this lesson in particular
is that due to the very novice level of the learners in this course
we have chosen to teach Git using a graphical user interface,
specifically the JupyterLab Git Extension \autocite{jl_git_ext},
as opposed to the Git command line tool.
Using a graphical user interface, and a newer one,
means that we need to more frequently update
and fix our lesson as the tool changes.
Command line tools in general tend to be more stable
(i.e., change less frequently)
compared to graphical user interfaces,
and are thus generally a more stable tool to build a lesson around.
However, the trade-off in this situation would be that
the Git command line tool is less intuitive for new learners,
especially those who are also new to the command line in general.

\hypertarget{lots-of-practice}{%
\subsection{Lots of practice}\label{lots-of-practice}}

The third pedagogy we argue for when teaching reproducibility is lots
and lots of practice.
Mastery of a subject often involves consolidating ideas, concepts
and theories into long-term memory,
which requires repetition \autocite{ebbinghaus1913memory}.
When we teach reproducibility topics,
we want students to also go a step further,
beyond knowledge of ideas, concepts and theories,
and induce a change in their behavior.
In other words
we want our students to form new habits
around how they perform data analyses - ones that are reproducible.

Habit formation can be defined as
the triggering of behavior from contextual cues \autocite{rebar_gardner_2019}.
In the context of reproducible workflows
these cues are the tasks that students desire to execute,
such as saving a file after adding new content
or wanting to share a document with a colleague.
Importantly,
habitual behavior has been proposed to protect individuals from motivational lapses,
where a desired good behavior is not expressed
due to a momentary lack of willpower \autocite{rebar_gardner_2019}.
By promoting the formation of habits in students,
we aim to change their behavior
so that they opt for the reproducible workflow ``by default'',
shielding them from relying on willpower to not ``take the easy way out''
and employ a familiar, but irreproducible workflow strategy.

In contrast to many other data science topics,
students often already have behavioral patterns in place
for reproducibility-related tasks,
such as how they organize and name their files,
and for how they collaborate with their colleagues.
However,
most often these practices
are not following a reproducible workflow
and can interfere with learning the new behaviours
we want our students to adopt.
When teaching reproducible workflows
we therefore need to replace the old behavior with a more reproducible version
as a response to the same cue.
Although it might sound like a complex task to unlearn an old behavior,
as well as learn a new one,
studies has shown that behavioral change is in fact facilitated
when substituting a desirable habit for an existing, undesirable one
rather than simply trying to unlearn the existing, undesirable habit
\autocite{evers_adriaanse_2011}.

Since habits are best learned through frequent, regular, and sustained cue exposure,
we argue that reproducibility requires more practice compared to other
data science topics.
To support the formation of reproducible workflow habits,
we therefore complement guided instruction
with plenty of embedded practice in the classroom,
where we intentionally pause during the demos
and say ``okay students your turn, do what i just did''.
This is a more controlled form of practice,
which sets students up for practicing these habits on their own.
As mentioned above,
we also provide students with worksheets
that can be used both in the classroom and at home,
as students receive feedback through automated software tests.

It is important to note that
habit formation is not a linear process.
Instead,
each successful action following a cue
adds to the formation of a new habit in an asymptotic fashion,
where the initial events are the most important
and the learning rate eventually plateaus as the habitual pattern solidifies.
While popular literature often refers to \textasciitilde30 days as ``all it takes''
to develop a new habit,
studies have reported that the median is at least around 70 days
before reaching the plateau phase of habit development \autocite{wardle_lally_2010,radel_fournier_2017}.
We therefore believe it is paramount
that learners have sustained frequent practice in reproducible workflows,
which is interleaved with other topics
where they would employ and benefit from these skills in real life.

To give students adequate time and practice to cement their reproducible workflow habits,
we have made intentional choices of which learning technologies and platforms
are used throughout the UBC Master of Data Science program
and when they are taught.
This ensures that students are practicing using reproducible tools
for the full ten months
as part of the course learning technology mechanics
(homework submission, grading, etc).
Students learn these basic reproducibility practices
in the first few weeks of class
and we interleave them as ``mechanical assignment requirements''
worth a small percentage of their grade
while completing assignments focused on other data science topics.
This strategy parallels how they will employ reproducibility
habits later in their careers
(a detailed example follows in the next section).
To challenge students who have more interest, experience, or skills in the area of reproducibility,
we routinely include demanding optional questions
in our homework assignments in the reproducibility courses
that allow these students to engage more deeply
with the material for a small amount of bonus points.
The feedback we have received
indicates that these strategies help to engage students
across the wide range of interest, experience
and skills in the area of reproducibility
that we see in our classrooms.

The decision to give students plenty of opportunities to practice reproducible habits
by adopting version control for student homework submission
was implemented across the Master of Data Science program from year one.
To make this feasible,
postdoctoral teaching and learning fellows were hired
to help faculty implement this in their courses
even if those faculty themselves were not familiar with version control.
The teaching team model in the Master of Data Science program
greatly facilitates us doing program-wide changes like this.
This stems from our shared vision for the program
(which was drafted collaboratively by the team),
our once-per term academic retreats
where we reflect on how things went in each course,
and our frequent communication and collaboration
at our weekly academic meetings.

\hypertarget{example-lessons-of-lots-of-practice}{%
\subsubsection{Example lesson(s) of lots of practice}\label{example-lessons-of-lots-of-practice}}

A specific example of how lots of practice
is used in almost all of the UBC Master of Data Science courses,
includes the use of version control,
particularly Git and GitHub,
as our course management system.
In these courses,
the homework instructions and assignments are distributed
to the students as GitHub repositories and the only way that they can submit
their homework is via their assigned GitHub repository.
Thus,
to complete and submit their assessments on any data science topic,
students must go through the cloning procedure
(or at least be able to somehow download their assignments from the GitHub website)
and upload their work back to their respective GitHub repository.
The tool we use to administer GitHub repositories in this way
is called \href{https://github.com/mgelbart/rhomboid}{\texttt{rhomboid}},
but there are other tools that can do this,
including \href{https://classroom.github.com/}{GitHub Classroom}
and \href{https://github.com/ubccpsc/classy}{Classy}.

To incentivize students to submit their assignment using Git
(as opposed to the GitHub web user interface)
we also assign some marks of each assignment (about 5\%) to a mechanics grade.
For this we assess whether they have at least three
commits associated with every single assignment
and have written meaningful commit messages.
By the end of this program the students have version controlled their
work in over 80 different GitHub repositories.
We hope this results in version control becoming a habit,
to the point that when they go to work on a project in the future
they will put the project under version control by default -
even when they leave the program
and are no longer receiving grades for doing this.
We choose tools such as Git and GitHub
when teaching version control
(and other data science reproducibility aspects)
since they are very popular in academia and industry -
meaning that students are likely to encounter them in their future work.
This sustained practice with real tools
not only enforces students' habits,
but also increases their proficiency using authentic reproducibility software
and when they run into problems,
it is in an environment
where they can easily reach out for help without feeling intimidated to ask.

One exciting technology that we have recently started incorporating
in our teaching of reproducible workflows is GitHub actions,
a tool for automating software workflows
that can be configured to be triggered by many different version control tasks,
including pushing new code to a GitHub repository.
This has allowed us
to automate the building of individual ``playgrounds'' of complex Git scenarios
that would take much effort and typically fail to stage in a large classroom.
One of the GitHub repositories that we created for these activities, called
\href{https://github.com/ttimbers/review-my-pull-request}{\texttt{review-my-pull-request}},
serves the purpose of providing a playground
where students can explore and practice
to learn how to use GitHub's code review feature for pull requests.
To use it,
students create their own copy of the repository on GitHub,
create a branch named \texttt{pr}
and then a pull request is automatically created for them by a bot.
After this quick and simple setup,
the students can spend the rest of the exercise
exploring how to perform code reviews on GitHub.

\hypertarget{conclusions}{%
\subsection{Conclusions}\label{conclusions}}

Over the past six years of teaching in the UBC Master of Data Science
and the UBC introductory undergraduate data science course,
we have identified the following key strategies
for effectively teaching reproducibility
in the data science classroom:
\emph{1)} providing extra emphasis on motivation
so that students understand why reproducibility is important and
buy into learning about it and practicing it,
\emph{2)} providing guided instruction so that it is not
a mystery of how you get started and what you need to do,
as well as \emph{3)} lots and lots of practice so that we
do not only teach them the ideas and the
concepts behind reproducibility, but so that students actually
change their data analysis habits and workflows into reproducible ones
that they can rely on after leaving the classroom.

\hypertarget{acknowledgements}{%
\subsection{Acknowledgements}\label{acknowledgements}}

We are grateful to the Master of Data Science and DSCI 100 teaching teams who helped shape
these opinionated practices for teaching reproducibility.

\hypertarget{declaration-of-interest-statement}{%
\subsection{Declaration of interest statement}\label{declaration-of-interest-statement}}

The authors declare no conflict of interest.

\newpage

\hypertarget{tables}{%
\subsection{Tables}\label{tables}}

\textbf{Table. 1}: Case studies that illustrate the significant real world consequences of reproducibility failures in data analyses.

\begin{longtable}[]{@{}lll@{}}
\toprule
\begin{minipage}[b]{0.34\columnwidth}\raggedright
Reproducibility error\strut
\end{minipage} & \begin{minipage}[b]{0.32\columnwidth}\raggedright
Consequence\strut
\end{minipage} & \begin{minipage}[b]{0.25\columnwidth}\raggedright
Source(s)\strut
\end{minipage}\tabularnewline
\midrule
\endhead
\begin{minipage}[t]{0.34\columnwidth}\raggedright
Limitations in Excel data formats\strut
\end{minipage} & \begin{minipage}[t]{0.32\columnwidth}\raggedright
Loss of 16,000 COVID case records in the UK\strut
\end{minipage} & \begin{minipage}[t]{0.25\columnwidth}\raggedright
\autocite{kelion_kelion_2020}\strut
\end{minipage}\tabularnewline
\begin{minipage}[t]{0.34\columnwidth}\raggedright
Automatic formatting in Excel\strut
\end{minipage} & \begin{minipage}[t]{0.32\columnwidth}\raggedright
Important genes disregarded in scientific studies\strut
\end{minipage} & \begin{minipage}[t]{0.25\columnwidth}\raggedright
\autocite{zeeberg2004mistaken,ziemann2016gene}\strut
\end{minipage}\tabularnewline
\begin{minipage}[t]{0.34\columnwidth}\raggedright
Deletion of a cell caused rows to shift\strut
\end{minipage} & \begin{minipage}[t]{0.32\columnwidth}\raggedright
Mix-up of which patient group received the treatment\strut
\end{minipage} & \begin{minipage}[t]{0.25\columnwidth}\raggedright
\autocite{wallensteen2018retraction}\strut
\end{minipage}\tabularnewline
\begin{minipage}[t]{0.34\columnwidth}\raggedright
Using binary instead of explanatory labels\strut
\end{minipage} & \begin{minipage}[t]{0.32\columnwidth}\raggedright
Mix-up of the intervention with the control group\strut
\end{minipage} & \begin{minipage}[t]{0.25\columnwidth}\raggedright
\autocite{wise_aboumatar_2019}\strut
\end{minipage}\tabularnewline
\begin{minipage}[t]{0.34\columnwidth}\raggedright
Using the same notation for missing data and zero values\strut
\end{minipage} & \begin{minipage}[t]{0.32\columnwidth}\raggedright
Paper retraction\strut
\end{minipage} & \begin{minipage}[t]{0.25\columnwidth}\raggedright
\autocite{turchin_whitehouse_2021}\strut
\end{minipage}\tabularnewline
\begin{minipage}[t]{0.34\columnwidth}\raggedright
Incorrectly copying data in a spreadsheet\strut
\end{minipage} & \begin{minipage}[t]{0.32\columnwidth}\raggedright
Delay in the opening of a hospital\strut
\end{minipage} & \begin{minipage}[t]{0.25\columnwidth}\raggedright
\autocite{picken_picken_2020}\strut
\end{minipage}\tabularnewline
\bottomrule
\end{longtable}

\newpage

\hypertarget{figures}{%
\subsection{Figures}\label{figures}}

\newpage

\hypertarget{figure-captions}{%
\subsection{Figure captions}\label{figure-captions}}

\begin{itemize}
\tightlist
\item
  \textbf{Fig 1:} Infamous comic from \href{https://xkcd.com}{xkcd.com} that highlights the difficulty of learning and using the version control software Git.
\item
  \textbf{Fig 2:} Latex errors are often cryptic to new learners.
\end{itemize}

\newpage
\singlespace
\footnotesize

\printbibliography

\end{document}